\documentclass[journal,twoside,web]{ieeecolor}
\usepackage{tmi}
\usepackage{cite}
\usepackage{amsmath,amssymb,amsfonts}
\usepackage{graphicx}
\usepackage{textcomp}
\usepackage{algorithm}
\usepackage{algpseudocode}
\usepackage{diagbox}

\def\BibTeX{{\rm B\kern-.05em{\sc i\kern-.025em b}\kern-.08em
    T\kern-.1667em\lower.7ex\hbox{E}\kern-.125emX}}
\begin{document}
\title{Segmentation-free PVC for Cardiac SPECT using a Densely-connected Multi-dimensional Dynamic Network}

\author{Huidong Xie, \IEEEmembership{Student Member, IEEE}, Zhao Liu, Luyao Shi, Kathleen Greco, Xiongchao Chen, \IEEEmembership{Student Member, IEEE}, Bo Zhou, \IEEEmembership{Student Member, IEEE}, Attila Feher, John C. Stendahl, Nabil Boutagy, \\ Tassos C. Kyriakides, Ge Wang, \IEEEmembership{Fellow, IEEE}, Albert J. Sinusas, and Chi Liu \IEEEmembership{Senior Member, IEEE}
\thanks{This work was supported by NIH grants R01HL154345, R01HL123949, T32HL098069, and S10RR025555.}
\thanks{Huidong Xie, Luyao Shi, Xiongchao Chen, Bo Zhou, Albert J. Sinusas, and Chi Liu are with the Department of Biomedical Engineering at Yale University.}
\thanks{Zhao Liu, Kathleen Greco, Albert J. Sinusas, and Chi Liu are with the Department of Radiology and Biomedical Imaging at Yale University}
\thanks{Attila Feher, John C. Stendahl, and Albert J. Sinusas are with the Department of Internal Medicine (Cardiology) at Yale University}
\thanks{Nabil Boutagy is with the Department of Pharmacology (Vascular Biology and Therapeutics) at Yale University}
\thanks{Tassos C. Kyriakides is with the Department of Biostatistics at Yale University.}
\thanks{Ge Wang is with the Department of Biomedical Engineering at Rensselaer Polytechnic Institute}
\thanks{Correspondence. Email: chi.liu@yale.edu }}
\maketitle

\begin{abstract}
In nuclear imaging, limited resolution causes partial volume effects (PVEs) that affect image sharpness and quantitative accuracy. Partial volume correction (PVC) methods incorporating high-resolution anatomical information from CT or MRI have been demonstrated to be effective. However, such anatomical-guided methods typically require tedious image registration and segmentation steps. Accurately segmented organ templates are also hard to obtain, particularly in cardiac SPECT imaging, due to the lack of hybrid SPECT/CT scanners with high-end CT and associated motion artifacts. Slight mis-registration/mis-segmentation would result in severe degradation in image quality after PVC. In this work, we develop a deep-learning-based method for fast cardiac SPECT PVC without anatomical information and associated organ segmentation. The proposed network involves a densely-connected multi-dimensional dynamic mechanism, allowing the convolutional kernels to be adapted based on the input images, even after the network is fully trained. Intramyocardial blood volume (IMBV) is introduced as an additional clinical-relevant loss function for network optimization. The proposed network demonstrated promising performance on 28 canine studies acquired on a GE Discovery NM/CT 570c dedicated cardiac SPECT scanner with a 64-slice CT using Technetium-99m-labeled red blood cells. This work showed that the proposed network with densely-connected dynamic mechanism produced superior results compared with the same network without such mechanism. Results also showed that the proposed network without anatomical information could produce images with statistically comparable IMBV measurements to the images generated by anatomical-guided PVC methods, which could be helpful in clinical translation.
\end{abstract}


\section{Introduction}
\label{sec:introduction}
\IEEEPARstart{S}{ingle} photon emission computed tomography (SPECT) is a nuclear imaging modality used to visualize and measure radiotracer activities within the body that is widely used for clinical purposes \cite{bockisch_hybrid_2009}. In SPECT imaging, the image quality is negatively affected by various factors, such as photon attenuation, photon scatterings, low photon sensitivities, and motion \cite{ritt_absolute_2011}. In addition to these, factors related to limited spatial resolution generally cause partial volume effects (PVEs) in which contributions from different tissues may be combined into a single voxel. Because different tissues have different patterns and kinetics of tracer uptake, PVE will cause undesirable blurring in the reconstructed images. In addition, the size of reconstructed voxels of SPECT is usually larger than that of other higher-resolution imaging modalities (such as computed tomography (CT)), leading to additional PVEs. In SPECT imaging, PVEs are usually represented as spill-over of photon counts between different tissues \cite{erlandsson_review_2012}. For example, in cardiac imaging, photons emitted from the myocardium may spill-over to the blood pool and vice versa, resulting in under-estimation or over-estimation of tracer activities in the reconstructed images.

The spatial resolution of SPECT imaging systems is characterized by their point spread functions (PSFs), and the goal of partial volume corrections (PVCs) is to reverse the effects of the system PSF. Using only the SPECT data, this can be simply done by deconvolution in the image domain or by incorporating PSF into the system matrix for iterative reconstruction \cite{reader_em_2003}. Nonetheless, both ways would result in undesirable artifacts due to loss of high-frequency information \cite{erlandsson_review_2012}. As discussed in \cite{strul_robustness_1999}, this issue can be effectively addressed by incorporating anatomical information for PVC. For cardiac SPECT imaging, our group has investigated a series of anatomical-guided PVC methods incorporating high-resolution contrast-enhanced CT angiography (CTA) images \cite{chan_noise_2016, liu_anatomical-based_2015}. Among these methods, we found that the iterative Yang (iY) \cite{yang_investigation_1996, erlandsson_review_2012} method is preferred, as this approach is robust to image noise \cite{mohy-ud-din_quantification_2018}. Despite its promising results, iY requires tedious image registration and segmentation steps between SPECT data and anatomical organ templates, which typically take about 10 hours per scan for the canine studies used in this work. iY also assumes perfect registration/segmentation between SPECT data and anatomical organ templates, which may not be practical in clinical settings due to motion artifacts, especially for cardiac imaging. Lastly, contrast-enhanced CT (CECT) is usually used to obtain anatomical organ templates, which introduces additional radiation exposure to patients along with the need of non-contrast CT for attenuation correction.

We previously proposed an atlas-based method to achieve segmentation-free PVC \cite{liu_fully_2017}. This method requires a set of atlas images and the corresponding segmentation templates. Non-rigid registrations were needed between the target CECT images and the atlas CECT images (target images refer to the test data in the context of a neural network). The transformation matrixes obtained from registrations were then applied to the atlas segmentation templates. Lastly, all the transformed segmentation templates were fused together for iY-PVC. Therefore, the manual segmentation step for iY-PVC can be avoided. However, this method is time and memory-consuming due to numerous non-rigid registration steps on high-resolution CECT image volumes. Also, its performance is highly dependent on the atlas database. The atlas-based method needs to be individually optimized for different species, patient populations, and even different sizes of canine studies. The paper showed that the atlas-based method led to high variances for the PVC results \cite{liu_fully_2017}. Lastly, the atlas-based method only skips the manual segmentation step, but contrast CT is still needed.

Deep learning represents a new class of reconstruction algorithm \cite{wang_perspective_2016} and may be an ideal tool to address the above-mentioned limitations of anatomical-guided or atlas-based PVC methods. In the past years, convolutional-based neural networks have been implemented in various medical imaging applications such as low-dose CT \cite{shan_3-d_2018}, few-view CT \cite{xie_deep_2019, xie_deep_2020}, low-dose SPECT \cite{aghakhan_olia_deep_2022}, few-view SPECT \cite{xie_increasing_2022, ryden_deep_2020}, attenuation map generations \cite{chen_direct_2022} etc. However, little attention has been given to deep-learning-based PVC.

All the proposed convolutional-based networks mentioned above were trying to learn static convolutional kernels throughout the networks. In the context of this paper, a convolutional kernel is defined as a set of filters in a convolutional layer. Recent studies showed that dynamic convolutional networks can significantly improve the performance for classification tasks \cite{li_omni-dimensional_2022}. Initially proposed by Chen \textit{et al.} \cite{chen_dynamic_2020} and Yang \textit{et al.} \cite{yang_condconv_2020}, in a dynamic convolutional layer, instead of applying the same convolutional kernel to all the input data after the network is fully trained, dynamic convolution learns a weighted combination of multiple convolutional kernels adapted based on the input features, with negligible extra computational cost. Recently, as pointed out by Li \textit{et al.} \cite{li_omni-dimensional_2022}, the previously proposed dynamic convolution limits the adaptive capability to only one dimension (i.e., the number of convolutional kernels), and other dimensions were omitted (i.e., kernel spatial size, input channel size, and output channel size). Li \textit{et al.} proposed an omni-dimensional dynamic convolution, which generalizes the dynamic mechanism to all the dimensions in convolutional networks. However, the dynamic weights of all the previously proposed dynamic mechanisms are calculated based on feature maps of only the previous layer. In this work, inspired by the Dense-net structure \cite{huang_densely_2017}, we proposed a densely-connected multi-dimensional dynamic (DC-Dy) mechanism, in which dynamic weights are obtained not only from the previous layer, but also all the preceding layers. This idea encourages the network to reuse feature maps from all the layers and improve adaptive capability of the dynamic mechanism. Compared with the onmi-dimensional dynamic mechanism, the proposed DC-Dy mechanism does not introduce additional parameters. To the best of our knowledge, dynamic convolutional networks have not been investigated so far in the field of medical imaging.

When training a medical image reconstruction network, global image quality metrics, such as mean-absolute-error (MAE) and structural similarity index measurement (SSIM), are widely implemented as the loss functions to optimize network parameters. Even though using these metrics has achieved promising results in various deep-learning networks, it is a well-known issue that these metrics do not necessarily reflect the true image quality in clinical settings, especially for cardiac images acquired on dedicated cardiac scanners like GE Discovery NM/CT 570c \cite{bocher_fast_2010} or D-SPECT \cite{gambhir_novel_2009}. Since these scanners have a field-of-view (FOV) smaller than the reconstructed matrix size, there are some insignificant features outside the FOV. But these relatively unimportant features share the same weight in MAE or SSIM calculations. In this work, we proposed to incorporate a clinically-relevant image quantification metric as part of the loss function for network optimizations. 

Coronary microvascular disease (CMVD) is a prevalent and critical global health problem, which is often unrecognized \cite{bradley_definition_2022}. Non-invasive methods to evaluate myocardial micro-circulatory function to accurately diagnose CMVD remains complicated and time consuming. We previously proposed a methodological framework for intramyocardial blood volume (IMBV) quantification as a metric for microvascular function obtainable with SPECT imaging of $^{99m}$Tc-labeled red blood cells (RBCs)\cite{mohy-ud-din_quantification_2018}. $^{99m}$Tc-RBCs is routinely used for cardiac blood-pool imaging in the assessment of regional and global left ventricular function. Previous results showed that SPECT images with anatomical-guided PVC methods produced more accurate estimation of IMBV\cite{mohy-ud-din_quantification_2018}. IMBV could served as a novel index to diagnose CMVD \cite{chung_non-invasive_2018}. The noninvasive diagnosis of CMVD is a challenging clinical problem especially in the presence of coexisting epicardial coronary artery disease \cite{feher_quantitative_2017}. Here, we focused on investigating $^{99m}$Tc-RBCs canine studies as an example for cardiac SPECT PVC, while expect the proposed method can be applied to other imaging tracers/modalities in the future. 

In this work, we proposed deep neural network with a densely-connected multi-dimensional dynamic (DC-Dy) mechanism for whole-volume cardiac SPECT PVC. Compared with the static counterpart, the proposed dynamic network demonstrated consistently better performance on canine studies acquired on the GE Discovery NM/CT 570c scanner for evaluation of IMBV. Additionally, based on the imaging tracer and medical application, IMBV was incorporated as a clinically-derived loss to optimize network parameters. Ablation experiments presented in this paper demonstrated the effectiveness of this clinically-derived loss function. Compared with the anatomical-guided iY method, the proposed deep-learning method demonstrated fast and comparable performance without CECT as prior knowledge. Lastly, we also tested the proposed network with both CECT and SPECT images as dual-channel input (but without segmented organ templates). The network trained with dual-channel input further enhanced the PVC quality without tedious image segmentation.

\section{Methodology}

\subsection{Data Acquisition}
A total of 28 resting canine studies scanned with $^{99m}$Tc-RBCs were included in this project to validate the effectiveness of the proposed deep-learning PVC method. Formulated in \eqref{eq1}, IMBV is defined as the ratio between the mean activities of the entire myocardium ($M_{myo}$) and the mean activities of the left ventricular blood pool ($M_{lv\text{-}blp}$). IMBV served as the primary metric in this work to evaluate the image quality and quantitative accuracy before and after PVC. After PVC, IMBV values are expected to decrease as the $M_{myo}$ decreases and $M_{lv\text{-}blp}$ increases by reducing the spill-over effects. $M_{myo}$ and $M_{lv\text{-}blp}$ were calculated using the manually-segmented 3D organ templates obtained from the CECT image volumes.

\begin{equation}
\mathrm{IMBV} = \frac{M_{myo}}{M_{lv\text{-}blp}}\label{eq1}
\end{equation}

Prior to imaging, about 3 ml arterial blood was withdrawn from each animal from a femoral artery catheter into a heparinized vacutainer. RBCs were then labeled in vitro with sodium pertechnetate. About 30 minutes after RBC labeling, $^{99m}$Tc-RBCs ($18.8\pm4.3 \mathrm{mCi}$) were intravenously injected into the imaging subjects. SPECT scans were then performed 15 minutes after the injections. The SPECT scan time was about 10 minutes. A low-dose non-contrast CT was performed afterward for attenuation correction. Then, CECT scans were performed with retrospective electrocardiogram (ECG) gating during end-expiration. The animals were mechanically ventilated and were under general anesthesia (1-2\% isoflurane and 55-60\% nitrous oxide) during the procedure \cite{feher_computed_2020}. The use of animal data in this study was approved by the Institutional Animal Care \& Use Committee (IACUC) of Yale University.

All the scans were performed on a GE Discovery NM/CT 570c dedicated cardiac SPECT/CT system. The SPECT scanner consists of 19 solid-state cadmium zinc telluride (CZT) detector modules to generate projections $P\in\mathbb{R}^{32\times32\times19}$ with pixel size $2.46\times2.46~\mathrm{mm^2}$. The system has pinhole collimators focusing on the heart with a FOV of about 19 $\mathrm{cm}$ in diameter. Images ($I\in\mathbb{R}^{70\times70\times50}$) were reconstructed using the maximum likelihood expected-maximization (MLEM) algorithm \cite{shepp_maximum_1982} with 80 iterations and voxel size $4\times4\times4~\mathrm{mm^3}$. No post-filtering was applied.

\subsection{Iterative Yang PVC}
The acquired SPECT list-mode data were rebinned into 8 cardiac gates during end-expiration to ensure alignments between CECT and SPECT. End-diastolic and end-systolic gates were selected for iY-PVC and network training/validation/testing. SPECT images were first resized to the dimension of CECT images. CECT images were then manually registered to SPECT images. 5 organ templates, including myocardium ($ROI_{myo}$), blood pool ($ROI_{blp}$), liver ($ROI_{liver}$), lung ($ROI_{lung}$), and background ($ROI_{bkg}$) were then generated by manual segmentations from contrast CT for iY-PVC. The generated templates are binary and cover the entire 3D volumes. iY-PVC was achieved by calculating voxel-wise correction factors through the process summarized in algorithm \ref{alg1}.

\begin{algorithm}

\caption{iY-PVC method}\label{alg1}
\begin{algorithmic}
\Require $I$ \Comment{Initial reconstructed image using MLEM}
\Require $S$ \Comment{The system matrix}
\Require $ROI_O$ \Comment{Segmented organ templates}\\
\hfill$O\in\{myo, blp, lung, liver, bkg\}$
\State $k \gets 0$
\State $I_k \gets I$

\While{$k<10$} \Comment{10 iterations for iY-PVC}
\State $O_{mean}\gets mean\{I_k(ROI_O)\}$ \Comment{Mean values in} \\
\hfill each organ
\State $T\gets \sum O_{mean} \cdot ROI_O$ \Comment{Noise-free template}
\State $T_{proj}\gets S\cdot T$ \Comment{Forward projection}
\State $T_{recon} \gets \mathrm{MLEM}(T_{proj})$ \Comment{MLEM reconstruction}
\State $F\gets T/T_{recon}$ \Comment{Correction factors}
\State $I_{k+1}\gets I\cdot F$ \Comment{Corrected image}
\State $k\gets k+1$

\EndWhile
\end{algorithmic}

\end{algorithm}

In iY-PVC, a noise-free organ template is generated first using the mean values within the segmented templates. The voxel-wise correction factors are equal to the ratio between the noise-free template and the MLEM-reconstructed template after forward-projection using the system matrix. This method typically requires a few iterations to converge. In this work, 10 iterations were used for iY-PVC. The manual registration took a few minutes per scan and the manual segmentation for five organs took around 10 hours per scan. The time-consuming process of generating accurate organ templates demonstrates the need for segmentation-free PVC method.

\subsection{Densely-connected Multi-dimensional Dynamic Convolution}

Mathematically, a static 3D convolutional operations \cite{yamashita_convolutional_2018} can be formulated as:

\begin{equation}
y = W\circledast x + B\label{eq2}
\end{equation}

where $x\in\mathbb{R}^{d\times w\times h \times C_{in}}$ and $y\in\mathbb{R}^{d\times w\times h \times C_{out}}$ represent input feature maps and output feature maps of the convolutional layer, respectively. $d$, $w$, and $h$ denote the spatial dimension of input/output feature maps, which may be different depending on the chosen parameters of the convolutional layer. $C_{in}$ and $C_{out}$ represent input and output channel dimensions. $W\in\mathbb{R}^{k\times k\times k \times C_{in} \times C_{out}}$ represents the weights in convolutional kernel. $k$ is the kernel spatial dimension. $B\in \mathbb{R}^{C_{out}}$ represents the bias term. $\circledast$ denotes the convolutional operations.

In the proposed DC-Dy convolutional layer, the kernel weight $W$ becomes adaptive based on the feature maps of all preceding layers on three kernel dimensions (i.e., spatial, input channel, and output channel dimensions). Specifically, three attention values, $a_{spa}\in\mathbb{R}^{k\times k\times k}$, $a_{in}\in\mathbb{R}^{C_{in}}$, and $a_{out}\in\mathbb{R}^{C_{out}}$, are obtained via an attention mechanism. $a_{spa}$, $a_{in}$, and $a_{out}$ are different for different input image volumes. Mathematically, \eqref{eq2} becomes:

\begin{equation}
y = (\frac{1}{3}W \odot (a_{spa} + a_{in} + a_{out})) \circledast x + B\label{eq3}
\end{equation}

The kernel weight $W$ is progressively weighted using $a_{spa}$, $a_{in}$, and $a_{out}$, and make it different for different input image volumes along with all spatial locations, input channels, and output channels. With this design, dynamic convolutional layer can capture more informative features than its static counterpart.

\begin{figure*}[!t]
\centerline{\includegraphics[width=\textwidth]{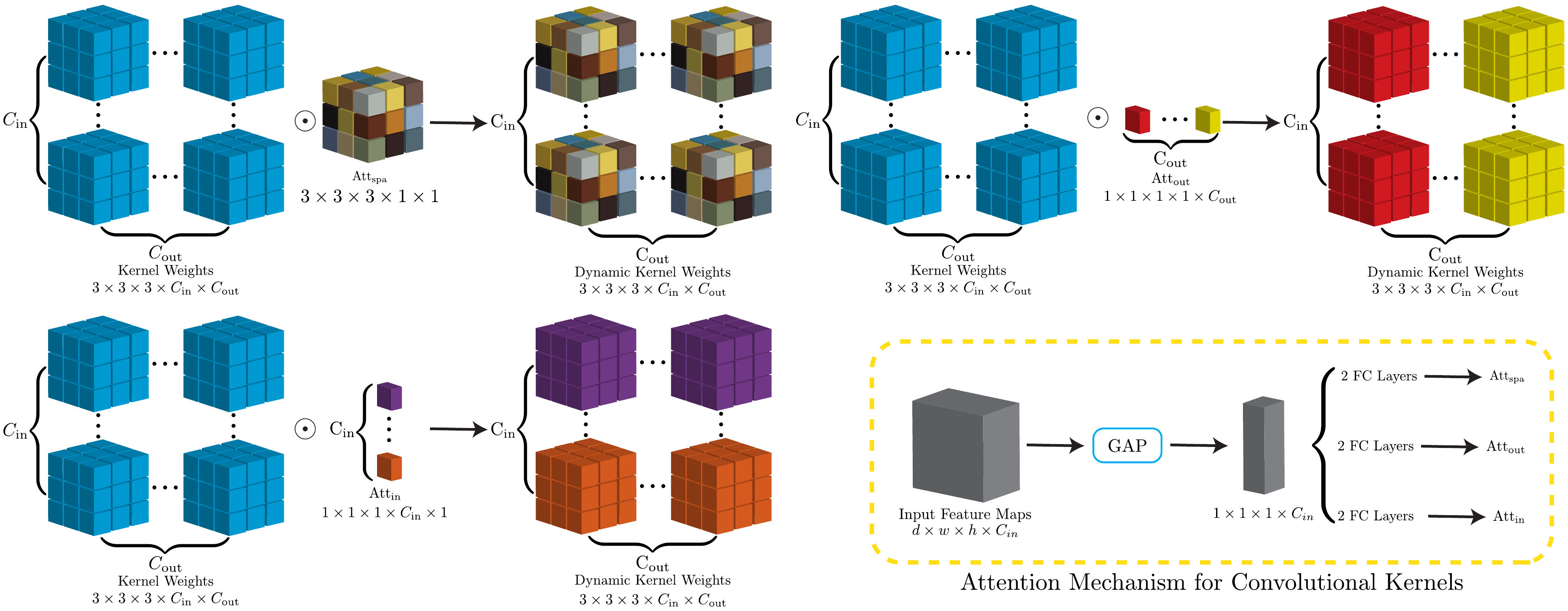}}
\caption{Graphical illustration of the proposed DC-Dy convolution with kernel size $3\times3\times3$, input channel $C_{in}$ and output channel $C_{out}$. The attention mechanism used for convolutional kernels is also presented here. Small light-blue cubes represent values in the static convolutional kernel (before applying the attention). Cubes with other colors represent how the attention values are applied. The dynamic kernel weights are obtained by multiplying the kernel weights and three attention values. Three dynamic kernel weights are then added together before performing the convolutional operations. Input feature maps are obtained using \eqref{eq4}.}
\label{fig1}
\end{figure*}

The attention mechanism used to obtain $a_{spa}$, $a_{in}$, and $a_{out}$ is adapted from the squeeze-and-excitation attention proposed by Hu \textit{et al.} \cite{hu_squeeze-and-excitation_2018}. The input features $x$ are first squeezed along spatial dimension using a global average pooling (GAP) operation ($\mathrm{GAP}(x)\in\mathbb{R}^{1\times1\times1\times C_{in}}$), followed by a fully connected (FC) layer and a rectified linear unit (ReLU) as the activation function. Another FC layer is used to project the output from the previous FC layer to the size of the attention values, followed by a sigmoid activation function to squeeze the attention values between 0 and 1. The 2 FC layers contain $2n$ and $n$ neurons, where $n=k^3$ for $a_{spa}$, $n=C_{in}$ for $a_{in}$, and $n=C_{out}$ for $a_{out}$. A graphical illustration of the proposed multi-dimensional dynamic convolutional with kernel size $3\times3\times3$ (i.e., $k=3$) is presented in Fig. \ref{fig1}. The attention mechanism used for the convolutional kernels is also presented in Fig. \ref{fig1}.

With proposed DC-Dy mechanism, three attention values are obtained based on input features of all preceding layers. Specifically, with the DC mechanism, the three attention weights (i.e., $a_{spa}$, $a_{in}$, and $a_{out}$) are obtained using feature maps from the previous layer ($(l-1)^{th}$) and all the preceding layers ($1^{st}$ to $(l-2)^{th}$ layers). Note that the DC mechanism only affects the three attention weights for convolutional kernels but not the input to the convolutional layers. Formulated in \eqref{eq4}, to avoid introducing additional parameters and memory burden, features for attention weights calculations are combined using additions:

\begin{equation}
x_l^{in} = \frac{1}{2}(x_{l-1} + x_{prev})\label{eq4}
\end{equation}

where $x_l^{in}$ and $x_l$ denote input features to the $l^{th}$ layer for attention weights calculations, and output features from the $l^{th}$, respectively. Note that by using \eqref{eq4}, features from the $(l-1)^{th}$ layer has the largest weight when calculating the three attention values for the $l^{th}$ layer, and weights of all preceding layers gradually diminish. Intuitively, features that are the closest to the $l^{th}$ layer should contribute most to the attention calculations. For the $1^{st}$ layer, the input image volume is used to calculate three attention values. Feature maps are resized for dimension matching before (\ref{eq4}).
\\

\subsection{Network Structure}
\begin{figure*}[!t]
\centerline{\includegraphics[width=\textwidth]{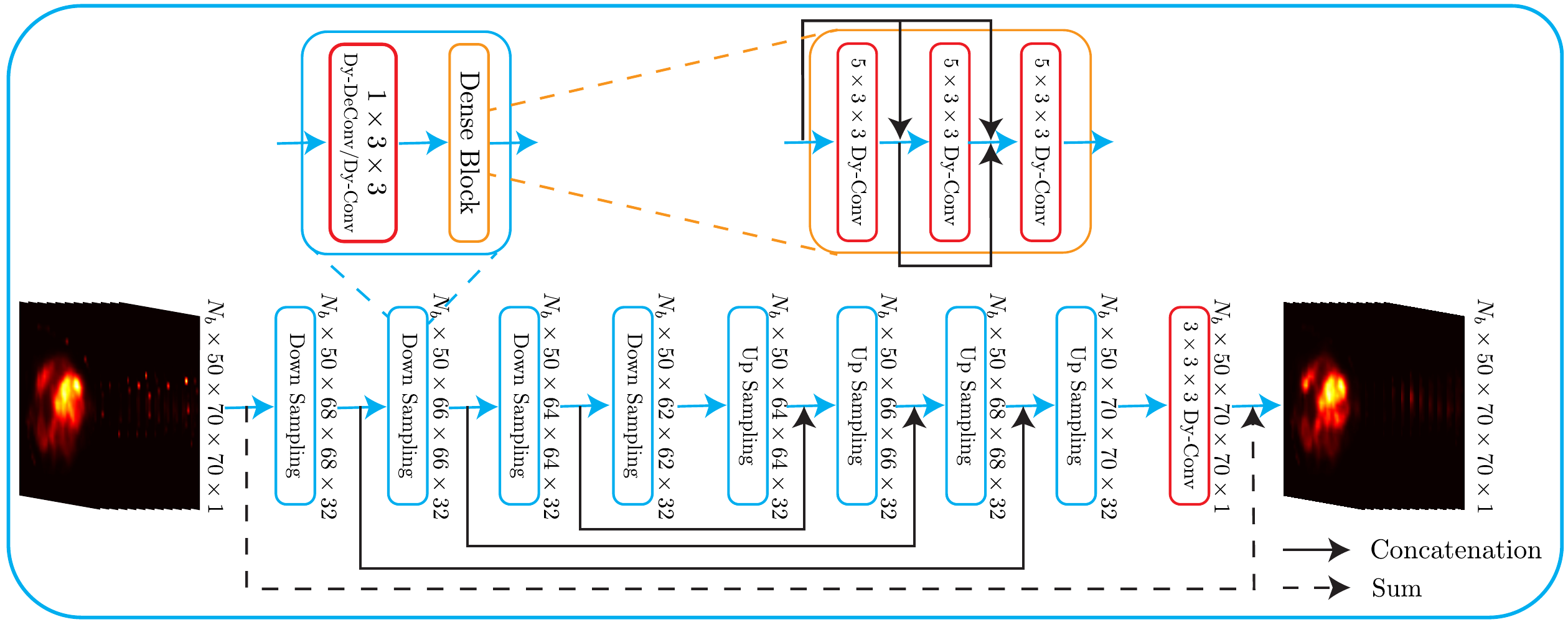}}
\caption{Overall structure of the proposed network. "Dy-Conv" and "Dy-DeConv" denote the proposed densely-connected multi-dimensional dynamic (DC-Dy) convolutional used for down-sampling and up-sampling respectively. Light-blue rectangles represent the down-sampling or up-sampling blocks. Red rectangles represent the "Dy-Conv" or "Dy-DeConv" operations. Orange rectangles represent the dense block. Numbers besides each block denotes the dimension of the layer output. $N_b$ represents the batch size.}
\label{fig2}
\end{figure*}

The proposed network adapts a U-net-like structure \cite{ronneberger_u-net_2015}, and it is presented in Fig. \ref{fig2}. The proposed network takes a batch of 3D non-PVC image volumes as input and tries to perform PVC without anatomical information from CECT. The dimensions for both input and output are $N_b\times50\times70\times70$, where $N_b$ represents the input batch size. As depicted in Fig. \ref{fig2}, the network contains 4 down-sampling blocks and 4 up-sampling blocks. Each of the down-sampling/up-sampling block has either a dynamic convolutional (Dy-Conv) or dynamic de-convolutional (Dy-DeConv) layer, followed by a dense-net block \cite{huang_densely_2017}. The size of dynamic convolutional kernel used for down-sampling or up-sampling is $1\times3\times3$ without zero-padding, so that the dimension along z-axis remains the same throughout the network. The dynamic convolutional kernel size used in dense-net block is $5\times3\times3$ with zero-padding to allow the network to capture contextual information for adjacent slices. ReLU activation functions were used after all the convolutional layers. Conveying paths are used to connect earlier layers and later layers. All convolutional layers contain 32 filters.

\subsection{Optimization and Training}
The proposed network was trained in a supervised manner with the MLEM-reconstructed images post-processed using iY as the reference. The objective function used to optimize all the trained networks in this study includes MAE and SSIM \cite{wang_image_2004}. MAE loss can be formulated as:

\begin{equation}
\ell_{\mathrm{MAE}}(Y,X)=\frac{1}{N_b W  H D}\sum_{i=1}^{N_b}||Y_i-X_i||_1,  \label{eqn:5}
\end{equation}

where $N_b$ is input batch size. $W=70$, $H=70$, and $D=50$ are the height, width, and depth of the input/output image volumes, respectively. $X$ and $Y$ represent output image volumes and the corresponding training label, respectively.

SSIM measures the structural similarity between two images, and it equals to 1 when two images are identical. In this work, the convolutional window used to calculate SSIM is set as $11\times11$. The SSIM formula is expressed as:

\begin{equation}
\mathrm{SSIM}(Y, X)=\frac{(2\mu_{Y}\mu_{X}+C_1)(2\sigma_{{YX}}+C_2)}{(\mu_{Y}^2+\mu_{X}^2+C_1)(\sigma_{Y}^2+\sigma_{X}^2+C_2)}  \label{eqn:6}
\end{equation}

where $C_1=(0.01\cdot R)^2$ and $C_2=(0.03\cdot R)^2$ are constants to stabilize the ratios when the denominator is too small. $R$ stands for the dynamic range of pixel values. $\mu_{Y}$, $\mu_{X}$, ${\sigma_{Y}}^2$, ${\sigma_{X}}^2$ and $\sigma_{{YX}}$ are the means of $Y$ and $X$, deviations of $Y$ and $X$, and the correlation between $Y$ and $X$ respectively. Since the network output and the training label are both 3D image volumes, we took advantage of that by including the SSIM values along 3 planes (transverse, coronal, and sagittal) into the objective function for network optimization. SSIM loss used to optimize the network parameters is expressed as: $\ell_{\mathrm{SSIM}}=1-SSIM(Y,X)$.

To emphasize the edges for different organs, the sobel operator (SO) \cite{Duda1973PatternCA} is used to obtain edge images from the network output and training label. The SO uses two separable filters to produce two gradient vector maps at each 2D spatial location. The MAEs between the gradient vector maps from network output and the training label are included as part of the objective function. Similar to the SSIM calculations, SO is applied for 3 planes, and the corresponding MAEs are included in the objective function. MAE loss calculated from the gradient vector maps can be expressed as:

\begin{equation}
\ell_{\mathrm{SO}}(Y, X)= \mathrm{MAE}(\mathrm{SO}(X)-\mathrm{SO}(Y)) \label{eqn:7}
\end{equation}

Since the GE Discovery NM Alcyone scanner used in this work has a FOV smaller than the reconstructed matrix, features inside and outside the FOV share the same weights in the SSIM and MAE calculations. This is not ideal because features inside the FOV should be more clinically relevant. As the anatomical information is available from CECT, and IMBV is a more relevant image quality metric before and after PVC, we took advantage of that by including the IMBV values into the objective function. IMBV loss function can be formulated as:

\begin{equation}
\ell_{\mathrm{IMBV}}(Y, X)= |\mathrm{IMBV}(X)-\mathrm{IMBV}(Y)| \label{eqn:8}
\end{equation}

IMBV values were calculated using the manually-segmented organ templates. Note that IMBV calculations were required only during the training stage. At the testing stage, no segmented template was needed.

SSIM values and MAE for gradient vector maps along 3 planes have the same weight in the composite loss function, which can be formulated as:

\begin{equation}
\begin{split}
\underset{{\theta_{Net}}}{\min}\,L = 
 \ell_{\mathrm{MAE}}(Y, X) + \lambda_a\,  \ell_{\mathrm{SSIM}}(Y,X) + \\ \lambda_b\, \ell_{\mathrm{SO}}(Y, X) + \lambda_c \ell_{\mathrm{IMBV}}(Y,X) 
 \label{eqn:9}
 \end{split}
\end{equation}

where $\lambda_a=0.8$, $\lambda_b=0.1$, $\lambda_c=0.1$ are hyper-parameters used to balance different loss functions. $\theta_{Net}$ represents all the trainable parameters in the network.

The Adam method \cite{DBLP:journals/corr/KingmaB14} was used to optimize all the trainable parameters in the network with 2 exponential decay rates $\beta_1=0.9$ and $\beta_2=0.999$. Xavier method \cite{pmlr-v9-glorot10a} was used to initialize all the convolutional kernel weights. All bias terms were initialized to 0. Among all the 28 canine studies, 15 were used for network training, 3 were used for validation, and the remaining 10 were used for network testing. Since end-diastolic and end-systolic gates are available for all the canine studies, there are a total of 30, 6, and 20 image volumes for network training, validation, and testing, respectively. Due to the limited amount of training data, 30 image volumes were augmented by rotating the image volume $30^\circ$ in the interval $(0^\circ,360^\circ)$ along all 3 planes, resulting in a total of $30\times11\times3+30=1,020$ image volumes for network training. Batch size $N_b=6$ was used.

\subsection{Evaluations}
In this work, image quality was quantitatively evaluated using SSIM, root-mean-squared-error (RMSE), peak signal-to-noise ratio (PSNR), and IMBV. One-way analysis of variance (ANOVA) and Tukey multiple comparison test \cite{keselman_tukey_1977} were used to evaluate the statistical significance in this study. In this work, p-value $p<0.05$ indicates statistical significance. Pairings were assumed in statistical testings. After PVC, the spill-over effects should be corrected, leading to a higher $M_{lv\text{-}blp}$ and lower $M_{myo}$. Hence, lower IMBV values typically represent better image quality in this work. The IMBV values obtained from iY-PVC results served as the gold standard for comparison.

Six networks were trained in this work to evaluate the effectiveness of all the network components and the proposed IMBV-derived loss function $\ell_{\mathrm{IMBV}}$. Six networks include: (1) a U-net (network depicted in Fig. \ref{fig2} without the dynamic mechanism); (2) a Dynamic U-net (U-net with the proposed multi-dimensional dynamic mechanism but without $\ell_{\mathrm{IMBV}}$, denoted as Dy-U); (3) a Dynamic U-net with CECT as the second-channel input but without $\ell_{\mathrm{IMBV}}$ (denoted as Dy-U-CT); (4) a Dynamic U-net with $\ell_{\mathrm{IMBV}}$ as an additional loss function (denoted as Dy-U-BV); (5) a Dynamic U-net with CECT and $\ell_{\mathrm{IMBV}}$ as an additional loss function (denoted as Dy-U-CT-BV); (6) a densely-connected dynamic U-net with $\ell_{\mathrm{IMBV}}$ as an additional loss function (denoted as DC-Dy-U-BV). Note that dynamic networks (2)-(5) did not implement the DC mechanism and the three attention values were obtained using features only from the previous layer. For networks with CECT as the second-channel input, CECT images were resized to SPECT dimension before concatenation.

It is expected that networks with dynamic mechanism would produce better results than the network without it, and networks with $\ell_{\mathrm{IMBV}}$ would produce images with better IMBV quantification than the networks without $\ell_{\mathrm{IMBV}}$. Images post-processed with iY served as the reference in this work.

\section{Results}

In this section, various network components are progressively added to demonstrate their effectiveness for cardiac SPECT PVC. Numbers are presented as $\mathrm{MEAN}\pm\mathrm{STD}$.

\subsection{Results on Dynamic U-net}

\begin{figure*}[!t]
\centerline{\includegraphics[width=\textwidth]{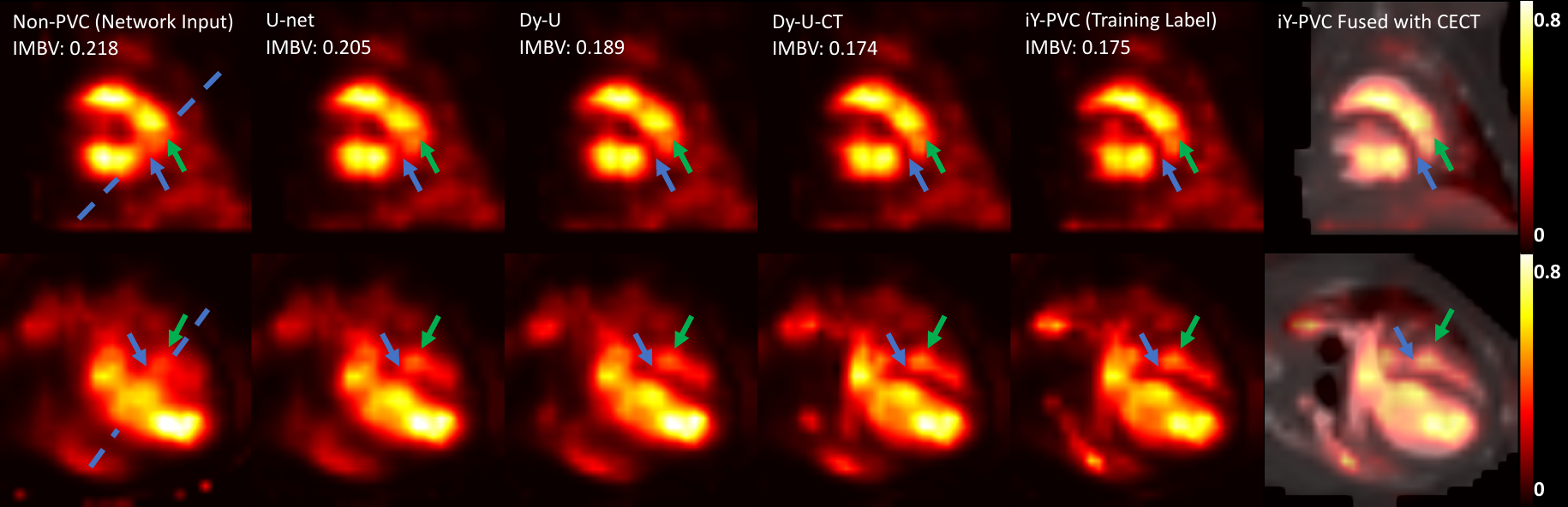}}
\caption{Transverse and coronal slices of a canine study reconstructed using different methods. Profile plots were generated along the dashed blue lines to present the partial volume effects and are presented in Fig. \ref{fig3_2}. Arrows with the same color point to the same regions in the images. Blue arrows point to regions that are over-estimated in the non-PVC images. Green arrows point to regions that Dy-U-CT performed better (closer to iY) than Dy-U. Corresponding IMBV values are also included in the figure.}
\label{fig3}
\end{figure*}

\begin{figure}[!t]
\centerline{\includegraphics[width=0.8\columnwidth]{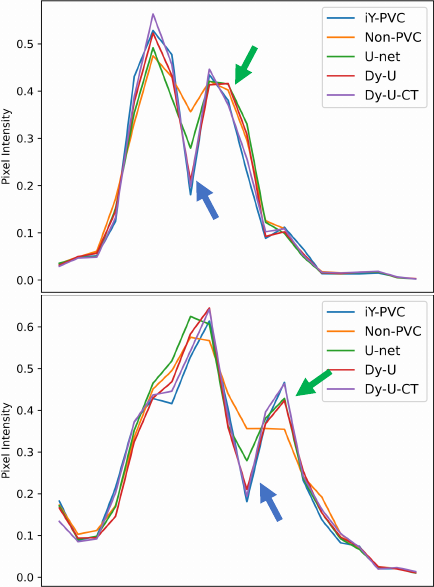}}
\caption{Profile plots along the dashed blue lines in Fig. \ref{fig3}. The upper and lower plots correspond to the images on the first and second rows of Fig. \ref{fig3}, respectively. Arrows with the same color point to the same regions in the images presented in Fig. \ref{fig3}.}
\label{fig3_2}
\end{figure}

In this sub-section, the first 3 networks (U-net, Dy-U, and Dy-U-CT) are compared. Both U-net and Dy-U used only the SPECT image data for PVC without CECT and segmented organ templates. Dy-U-CT concatenated CECT and SPECT data together as input without segmented organ templates. Since CECT served as the prior knowledge in the iY-PVC method, we expected Dy-U-CT should capture more anatomical information from CECT and perform better than the networks without CECT. 

One sample canine study was selected and presented in Fig. \ref{fig3}. All three deep learning methods produced images with reduced PVEs. Especially in the coronal slice, the contour of the left-ventricular blood-pool is fused with the myocardium region and is not clearly visible due to spill-over effects. The deep learning methods significantly improved reconstruction results, and the contours of each organ became more aligned with the CECT image. In addition to visual observations in the figure, the lower IMBV values and the profile plots (Fig. \ref{fig3_2}) demonstrated that networks with dynamic mechanisms produced even better results than the U-net. As indicated by the blue arrows in the transverse slice in Fig. \ref{fig3}, U-net did not perform well at the septal wall between the right and left ventricular blood-pools. Using the same network design with the proposed multi-dimensional dynamic mechanism, Dy-U demonstrated superior performance over U-net. Dy-U-CT produced images with better estimations of the blood-pool region than Dy-U did (green arrows in the transverse slice of Fig. \ref{fig3}). For this study, images reconstructed using Dy-U-CT have lower IMBV values than iY-PVC images, even without segmented organs.

The average IMBV values across all the 20 testing canine studies are $0.209\pm0.042$, $0.193\pm0.038$, $0.185\pm0.037$, $0.172\pm0.033$, and $0.167\pm0.033$ for Non-PVC, U-net, Dy-U, Dy-U-CT, and iY-PVC results, respectively. This downward trend is consistent with our expectation. Statistically significant differences were observed between all 5 groups ($p<0.001$), except for the difference between IMBV values obtained from iY-PVC and Dy-U-CT, which had a p-value $p=0.152$. This p-value demonstrates the effectiveness of using CECT as network input. Registered CECT images fused with the iY-PVC SPECT images are also included in Fig. \ref{fig3}.

\subsection{Results on Dynamic U-net with IMBV Loss Function}
\begin{figure*}[!t]
\centerline{\includegraphics[width=\textwidth]{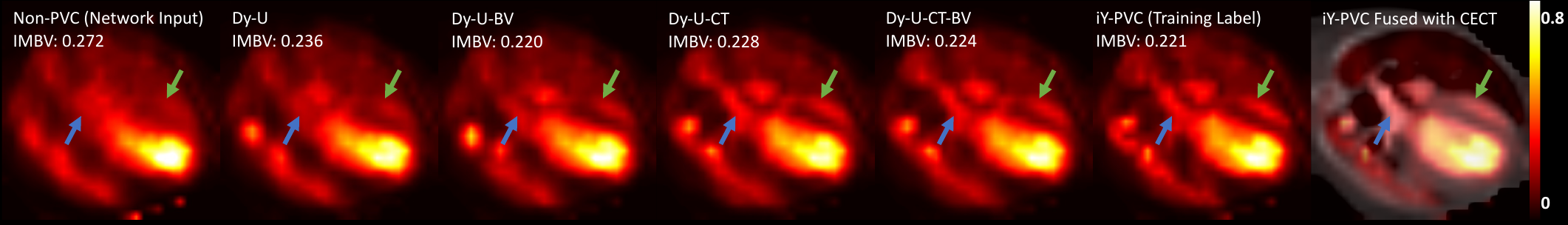}}
\caption{Coronal slices of a canine study reconstructed using different methods. Blue arrows point to anatomical features of the heart. Corresponding IMBV values are also included in the figure.}
\label{fig4}
\end{figure*}

The Dy-U and Dy-U-CT were re-trained with the $\ell_{\mathrm{IMBV}}$ (denoted as Dy-U-BV, and Dy-U-CT-BV, respectively) as an additional loss function to demonstrate the effectiveness of the proposed IMBV-derived loss function $\ell_{\mathrm{IMBV}}$. Ideally, using $\ell_{\mathrm{IMBV}}$ would force the network to concentrate more on the cardiac region and produce images with IMBV measurements closer to those of iY-PVC images than images generated by the networks without $\ell_{\mathrm{IMBV}}$. With the additional $\ell_{\mathrm{IMBV}}$, the average IMBV values of the 20 testing canine studies were $0.174\pm0.036$ and $0.166\pm0.033$ for Dy-U-BV and Dy-U-CT-BV, respectively. Compared with iY-PVC, the p-values were $p=0.009$ and $p=0.986$ for Dy-U-BV and Dy-U-CT-BV, respectively. Another canine study was selected and presented in Fig. \ref{fig4}. For the SPECT-only networks, the network with $\ell_{\mathrm{IMBV}}$ (Dy-U-BV) generated images with better reconstructed anatomical structure of the heart (blue arrows in Fig. \ref{fig4}), and less PVEs (green arrows in Fig. \ref{fig4}), as compared with Dy-U. For the networks using CECT as a prior knowledge (Dy-U-CT and Dy-U-CT-BV), despite the improved quantitative accuracy, they produced visually similar reconstructions.

\subsection{Results on Densely-connected Dynamic U-net}
\begin{figure*}[!t]
\centerline{\includegraphics[width=\textwidth]{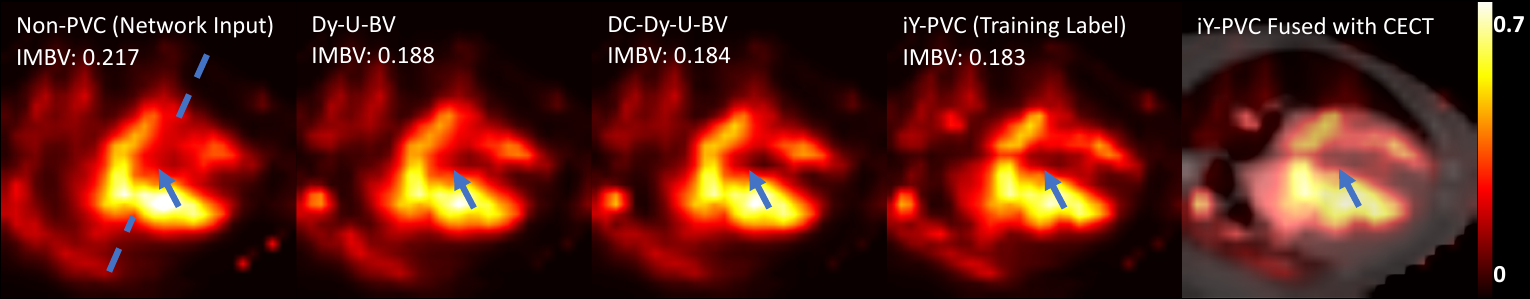}}
\caption{Coronal slices of a canine study reconstructed using different methods. Profile plot was generated along the dashed blue lines to present the partial volume effects and is presented in Fig. \ref{fig5_2}. Blue arrows point to regions with noticeable spill-over effects. Corresponding IMBV values are also included in the figure.}
\label{fig5}
\end{figure*}

The proposed Dy-U-CT-BV network already produces images with no statistically significant different IMBV quantification using the multi-dimensional dynamic mechanism and $\ell_{\mathrm{IMBV}}$ but without the DC mechanism. However, most available GE Alcyone scanners installed globally do not have an integrated CT system, and an additional CT scan also results in higher radiation dose to patients. Hence, a SPECT-only network has a higher potential of clinical translation. With the proposed DC-Dy mechanism and the $\ell_{\mathrm{IMBV}}$, the DC-Dy-U-BV network also produced images with no statistically significant IMBV measurements to the iY method ($0.169\pm0.034, p=0.946$) using only the SPECT data. On the other hand, without the DC mechanism, the average IMBV values produced by the Dy-U-BV network is $0.174\pm0.036$ with $p=0.009$ when compared with iY. One sample canine study was selected and presented in Fig. \ref{fig5}. The network with the DC-Dy mechanism produced images that are more aligned with the iY images.
\\
\\

\begin{figure}[!t]
\centerline{\includegraphics[width=0.8\columnwidth]{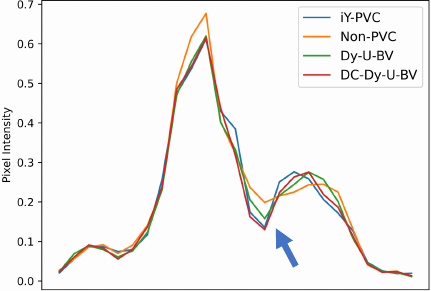}}
\caption{Profile plot along the dashed blue lines in Fig. \ref{fig5}. Blue arrows point to the same regions in the images presented in Fig. \ref{fig5}.}
\label{fig5_2}
\end{figure}

\subsection{Additional Quantitative Evaluations}
\begin{figure*}[!t]
\centerline{\includegraphics[width=\textwidth]{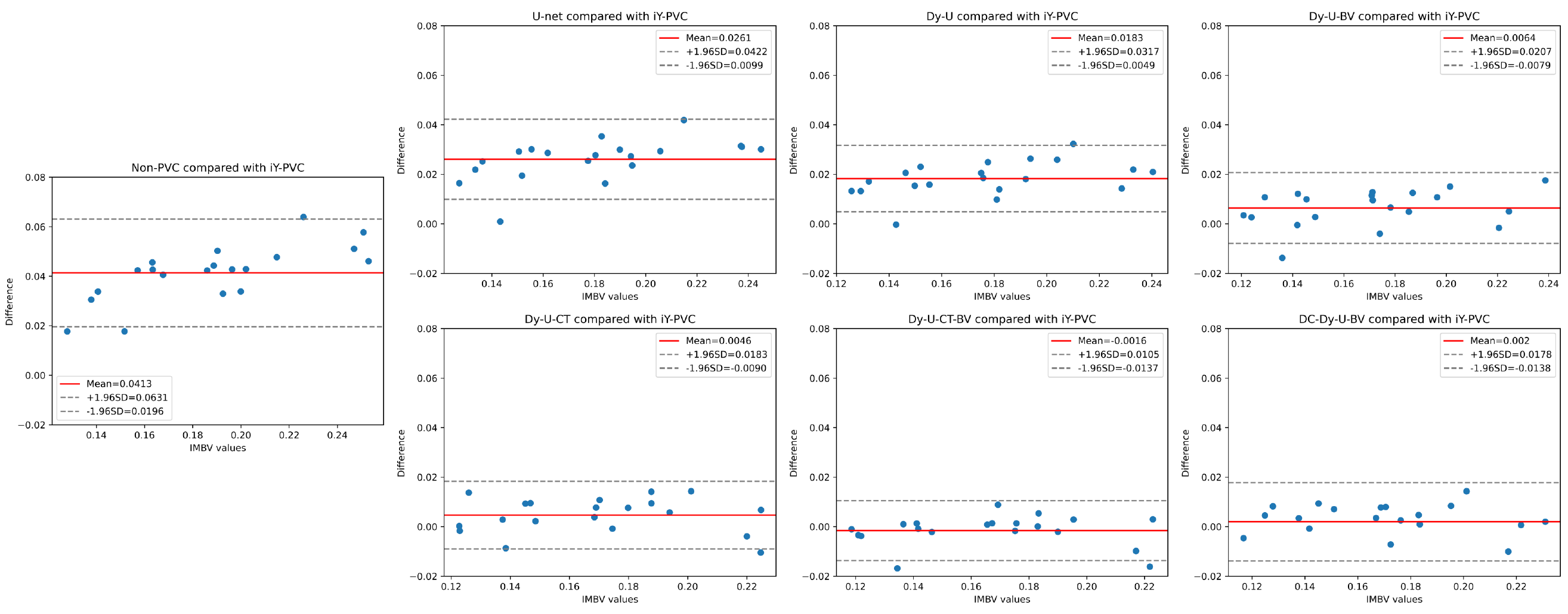}}
\caption{Bland-Altman plots for comparing non-PVC and deep learning methods. Results obtained from iY-PVC were used as the reference. IMBV values were used as the data points in the plots.}
\label{bland_altman}
\end{figure*}

\begin{figure*}[!t]
\centerline{\includegraphics[width=\textwidth]{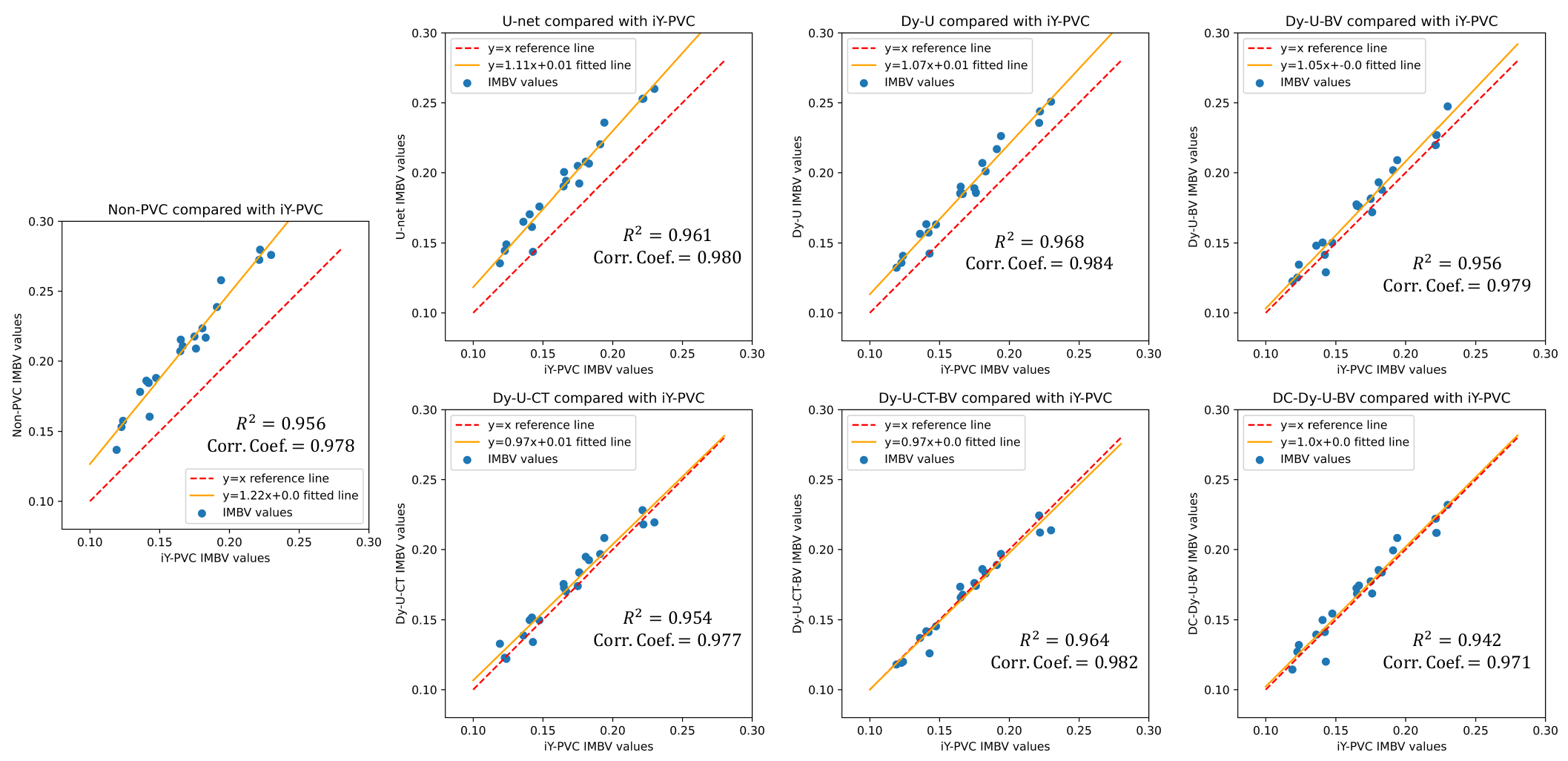}}
\caption{Linear fitting plots for comparing non-PVC and deep learning methods. Results obtained from iY-PVC were used as the reference. IMBV values were used as the data points in the plots. Coefficient of determination ($R^2$) and Pearson correlation coefficient (Corr. Coef.) are included in the linear fitting plots.}
\label{linear_fitting}
\end{figure*}

In order to compare results across all the 20 testing canine studies, 7 Bland-Altman plots and 7 linear fitting plots were generated using the IMBV values obtained from all the testing canine studies and are presented in Fig. \ref{bland_altman}-\ref{linear_fitting}, using the values obtained from iY-PVC results as the reference. The effectiveness of different network components can be clearly observed in Fig. \ref{bland_altman}-\ref{linear_fitting}.

\begin{figure}[!t]
\centerline{\includegraphics[width=\columnwidth]{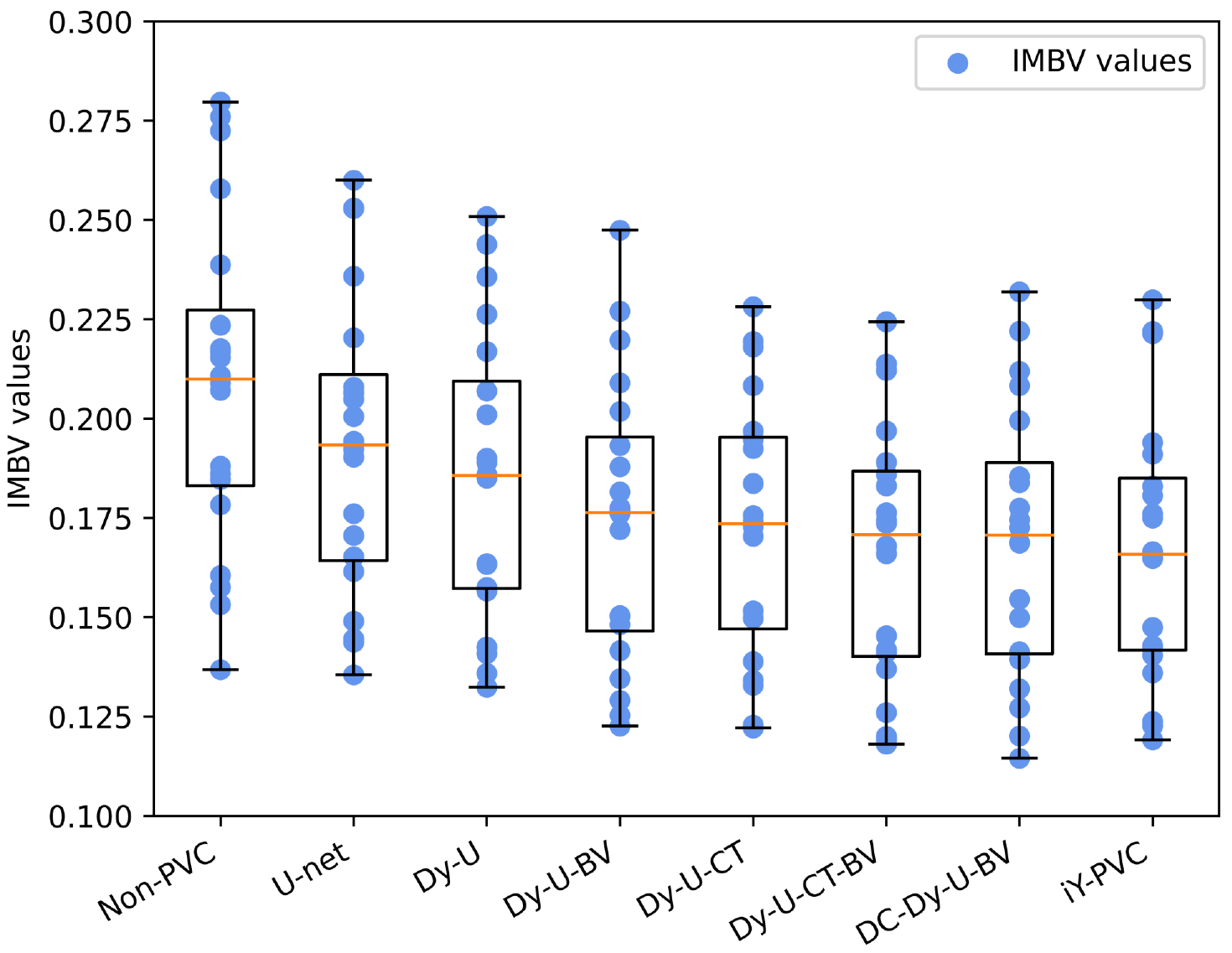}}
\caption{A box plot summarizing IMBV values obtained from all the methods}
\label{box_plot}
\end{figure}

A box plot summarizing the IMBV values of all the networks is also presented in Fig. \ref{box_plot}. After applying the $\ell_{\mathrm{IMBV}}$ and using CECT as the second-channel input, no statistical difference was observed for IMBV quantification between Dy-U-CT-BV results and the iY-PVC results ($p=0.986$). Without CECT as network input, the p-value between Dy-U-BV and iY-PVC results was $p=0.0089$. The difference between Dy-U-CT and iY-PVC was also not significant ($p=0.153$). With only the SPECT data, network with the proposed DC-Dy mechanism and $\ell_{\mathrm{IMBV}}$ (DC-Dy-U-BV) also results similar IMBV quantification to iY ($0.169\pm0.034, p=0.946$).

In addition to IMBV calculations, other image quality measurements, including SSIM, PSNR, and RMSE are presented in Table \ref{table1}. Using the same loss function and training strategy, Dy-U benefited from the proposed dynamic convolution and outperformed the U-net with superior quantitative measurements ($p=0.003$, $p=0.018$, $p=0.039$, and $p<0.0001$ for SSIM, RMSE, PSRN, and IMBV, respectively).

Dy-U-CT, the network using CECT as additional input, produced results with better quantitative results than Dy-U did ($p<0.0001$ for SSIM, RMSE, PSNR, and IMBV).

With $\ell_{\mathrm{IMBV}}$ as an additional loss function, both Dy-U-BV and Dy-U-CT-BV produced images with more accurate IMBV measurements compared with the networks without $\ell_{\mathrm{IMBV}}$ ($p<0.0001$ for both networks when compared with their counterparts).

As presented in Table \ref{table1}, compared with the networks without $\ell_{\mathrm{IMBV}}$, networks with $\ell_{\mathrm{IMBV}}$ produced images with superior IMBV measurements but similar SSIM, PSNR, and RMSE values. We believe such a small difference in the measured values will not affect overall image quality, especially in clinical settings. The DC mechanism also resulted in similar SSIM, PSNR, and RMSE values compared with its counterpart (Dy-U-BV) but superior IMBV measurements. We believe the improvements in IMBV quantification are more meaningful in clinical settings, especially for the problem targeted in this work.

\begin{table*}[!h]
\centering
\caption{Quantitative assessment on different methods ($\mathrm{MEAN}\pm\mathrm{STD}$). For each metric, the best result is marked in \textcolor{red}{red}, and the second best result is marked in \textcolor{blue}{blue}. The measurements were obtained by averaging the values on the 20 testing canine studies. Slices that do not contain the heart are excluded from calculations. For IMBV calculations, numbers that are closest to the iY-PVC results are considered the best.}
\resizebox{\textwidth}{!}{
\begin{tabular}{|c|c|c|c|c|c|c|c|c|}
\hline
     &  Non-PVC         &   U-net  &    Dy-U     &  Dy-U-BV & DC-Dy-U-BV & Dy-U-CT      & Dy-U-CT-BV & iY-PVC \\
\hline
IMBV  & { }{ }$0.209\pm0.042$ & { }{ }$0.193\pm0.038$  & { }{ }$0.185\pm0.037$  & $0.174\pm0.036$& { }{ }\textcolor{blue}{$0.169\pm0.034$}  & { }{ }$0.172\pm0.033$  & { }{ }$\textcolor{red}{0.166\pm0.033}$    & { }{ }$0.167\pm0.033$ (Reference)\\
\hline
PSNR & $35.552\pm2.095$ & $37.130\pm2.485$ & $37.641\pm2.803$ & $37.686\pm2.642$ & $37.748\pm2.804$ &\textcolor{blue}{$40.319\pm3.598 $} & \textcolor{red}{$40.434\pm3.662$}   & \diagbox{}{} \\
\hline
SSIM & { }{ }$0.967\pm 0.016$ & { }{ }$0.979\pm0.014$  & { }{ }$0.9822\pm0.0155$  & $0.983\pm0.015$& { }{ }$0.9824\pm0.015$  &{ }{ }\textcolor{blue}{$0.9873\pm0.0166$}  & { }{ }\textcolor{red}{$0.9875\pm0.0166$}    & \diagbox{}{}\\
\hline
RMSE  & { }{ }$0.017\pm0.006$ & { }{ }$0.015\pm0.006$  & { }{ }$0.0139\pm0.0062$  & $0.0138\pm0.0060$& { }{ }$0.0137\pm0.0061$  & { }{ }\textcolor{blue}{$0.0107\pm0.0067$}  & { }{ }\textcolor{red}{$0.0106\pm0.0067$}    & \diagbox{}{}\\
\hline
\end{tabular}
}
\label{table1}
\end{table*}

\subsection{Additional Ablation Studies}
Two additional ablated networks were trained to demonstrate the effectiveness of $\mathrm{IMBV}$ as part of the loss function and CECT as a second-channel input to the network. These two networks are denoted as U-net-BV (U-net with $\mathrm{IMBV}$ as an additional loss function) and U-net-CT (U-net with CECT as a second-channel input). Corresponding quantitative measurements are presented in Table \ref{table2}.

\begin{table}[!h]
\centering
\caption{Quantitative assessment on different methods ($\mathrm{MEAN}\pm\mathrm{STD}$). The measurements were obtained by averaging the values on the 20 testing canine studies. Slices that do not contain the heart are excluded from calculations.}
\resizebox{0.8\columnwidth}{!}{
\begin{tabular}{|c|c|c|}
\hline
     &  U-net-BV         &   U-net-CT  \\
\hline
IMBV  & { }{ }$0.180\pm0.034$ & { }{ }$0.177\pm0.033$  \\
\hline
PSNR & $36.903\pm2.383$ & $38.980\pm3.214$  \\
\hline
SSIM & { }{ }$0.980\pm 0.014$ & { }{ }$0.982\pm0.015$  \\
\hline
RMSE  & { }{ }$0.015\pm0.006$ & { }{ }$0.012\pm0.006$  \\
\hline
\end{tabular}
}
\label{table2}
\end{table}

Similar to the dynamic networks, with $\mathrm{IMBV}$ as an additional loss function, U-net-BV produced images with superior IMBV quantification ($p<0.0001$) but with similar SSIM, PSNR, and RMSE values, compared with the U-net.

With CECT images as additional prior knowledge, U-net-CT produced images with superior quantitative measurements compared with U-net ($p<0.02$ for SSIM, RMSE, PSNR, and IMBV). However, its dynamic counterpart (Dy-U-CT) still outperformed U-net-CT across all the chosen metrics ($p<0.05$ for SSIM, RMSE, PSNR, and IMBV), which again demonstrated the effectiveness of dynamic convolutions in the case of cardiac SPECT PVC.

\section{Discussion}
Ischemic heart disease (IHD) is considered a leading cause of mortality globally, accounting for more than 9 million deaths in 2016 \cite{nowbar_mortality_2019}. IHD can be divided into obstructive coronary artery disease (CAD) and/or CMVD. CAD affects larger epicardial arteries ($>$500 $\mathrm{\mu m}$), while CMVD affects smaller ones ($<$500 $\mathrm{\mu m}$). CMVD exists in a large proportion of patients with IHD and/or other cardiovascular diseases, and it may or may not co-exist with CAD \cite{bradley_definition_2022}. IMBV could serve as a novel index for micro-vascular function to diagnose CMVD independent of CAD. Our previous work \cite{mohy-ud-din_quantification_2018} showed that SPECT imaging using $^{99m}$Tc-RBCs could be implemented as a non-invasive imaging technique for IMBV quantification. However, physical limitations of SPECT imaging result in PVEs and compromise the quantitative accuracy. We previously demonstrated that SPECT images reconstructed with iY significantly improved IMBV quantification \cite{mohy-ud-din_quantification_2018}. 

However, iY is not practical to be implemented in reality due to tedious image registration and segmentation steps. Additional radiation dose introduced from CT scans with or without contrast administration is also not ideal in human studies. In this work, we proposed a deep-learning technique to perform fast and consistent PVC without requiring segmented organ templates from CECT. The proposed network is featured with a U-net-like structure, the DC-Dy convolution, and an IMBV-derived loss function. Conventional convolutional-based networks aim to learn static convolutional kernels throughout the network. After training, the same set of convolutional kernels is applied to all the testing images, which is not ideal due to the large variability between imaging subjects. In this work, the proposed DC-Dy convolution allows the network to have adaptive convolutional kernels based on the input data to capture more informative contextual features for better performance and compensate for the variances between imaging subjects. Specifically, the DC-Dy mechanism in the network aims to assign different weights to the convolutional kernels along different dimensions for each input image volume. By doing so, the convolutional kernels in the proposed network can be adjusted even after the network is fully trained. Compared with the recently proposed onmi-dimensional dynamic convolution \cite{li_omni-dimensional_2022}, the proposed DC-Dy mechanism allows the network to learn the attention values based on features not only from the previous layer, but also all the preceding layers, without introducing additional parameters. This newly proposed mechanism further strengthens the adaptive capability of the network and improves IMBV quantification. To the best of our knowledge, this is the first work to investigate dynamic network in the medical imaging field.

The proposed network with both SPECT and CECT (without segmented organ templates) was validated with 28 canine studies, and produced images with comparable IMBV measurements to the iY-PVC images ($p=0.986$). Since most installed GE Alcyone scanners are SPECT-only systems, the use of co-registered CECT images is not as easily accomplished. We also demonstrated that using only the SPECT data, the network with DC-Dy mechanism and $\ell_{\mathrm{IMBV}}$ (DC-Dy-U-BV) produced images with comparable IMBV measurements to the iY-PVC images ($p=0.946$).

\begin{figure}[!t]
\centerline{\includegraphics[width=\columnwidth]{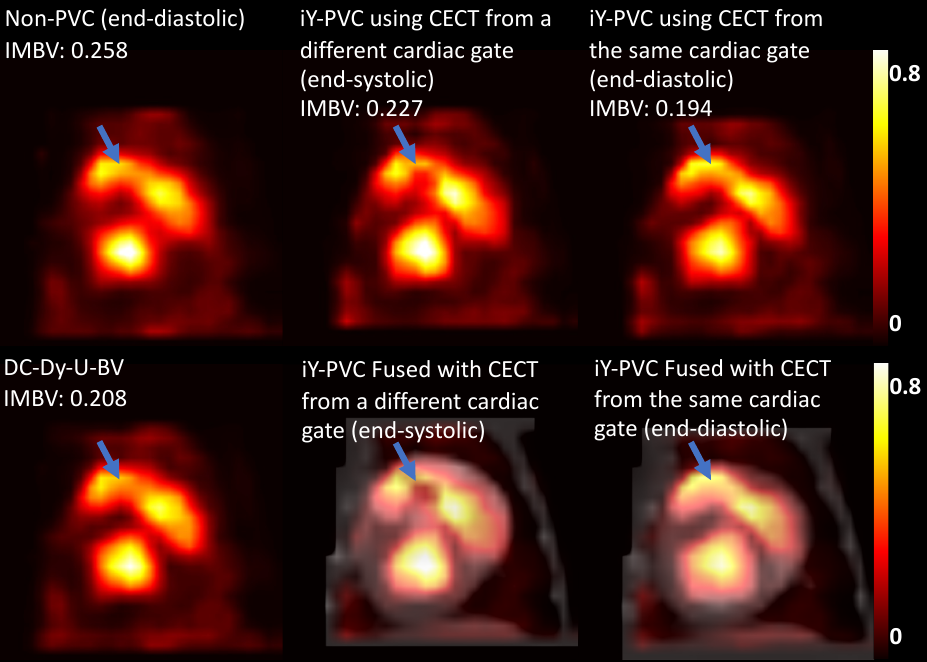}}
\caption{A transverse slice of a canine study reconstructed using iY with CECT from a different cardiac gate to demonstrate the limitations of iY. Blue arrows point to undesired artifacts introduced by imperfect registration between CECT and SPECT.}
\label{fig7}
\end{figure}

Note that iY-PVC method assumes perfect alignments between SPECT images and segmented organ masks. In this work, to ensure perfect alignments, CECT images were acquired with retrospective ECG gating during end-expiration. Such procedure is slow and will introduce additional radiation. To demonstrate that the image quality degrades severely in the case of imperfect registration, iY-PVC was performed on a canine study (acquired in end-diastolic gate) using CECT from another cardiac gate (end-systolic gate). One transverse slice was selected and presented in Fig. \ref{fig7}. As pointed out by the blue arrows in Fig. \ref{fig7}, if the CECT is not well-aligned with the SPECT images, undesired artifacts were introduced in the reconstructed images. The proposed deep-learning network overcame this limitation as CECT and segmented organ templates were not required. As indicated by the IMBV values, imperfectly registered CECT also negatively affects quantitative accuracy. With only the SPECT data as input, the proposed network demonstrated the potential of accurate IMBV quantification without CECT information and alleviate the concern of any SPECT-CT mismatch.

 In this work, the images acquired with $^{99m}$Tc-RBCs are used for IMBV quantification. However, superior SSIM or MAE values do not directly relate to better IMBV measurements. We proposed to incorporate the IMBV-derived metric into the overall loss function for network training. The network can then concentrate more on the cardiac region to produce images with more accurate IMBV quantification for potentially better clinical results. It is worth noting that dynamic network with $\ell_{\mathrm{IMBV}}$ (Dy-U-BV) has similar performance to the dynamic network with CECT (Dy-U-CT) in terms of IMBV quantification, which demonstrated the effectiveness of $\ell_{\mathrm{IMBV}}$. We believe that similar ideas incorporating the clinical quantification measurements into loss functions could be implemented for a wide range of clinical problems and/or other imaging modalities/tracers. 

The proposed multi-dimension dynamic convolution is not limited to PVC for cardiac SPECT. Given its dynamic nature, the proposed dynamic mechanism could be implemented in various neural networks for medical image reconstruction for different imaging modalities to achieve better performance.

The proposed loss function $\ell_{\mathrm{IMBV}}$ and the DC mechanism produced images with statistically better IMBV quantification but did not lead to noticeable improvements in SSIM, RMSE, and PSNR measurements. We suspect that it may be attributed to the fact that IMBV calculations only focus on the heart, while other metrics are based on the entire image volumes. Also, SSIM, RMSE, and PSNR were calculated using the iY-PVC images as the gold-standard. However, iY is not a perfect method for PVC as it relies on numerous assumptions \cite{erlandsson_review_2012}. Thus, we believe IMBV is a more relevant metric in this work to evaluate the image quality before and after PVC.

The GE Discovery NM/CT570c and 530c scanners are equipped with 19 CZT detectors mounted on an L-shape arc that covers a nearly $180^\circ$ range for stationary imaging. Therefore, these scanners essentially represent limited-view imaging systems with truncated projections. Our previous work \cite{xie_increasing_2022} proposed a multi-angle reconstruction approach to acquire multi-angle projections for improved image quality with detector gantry rotations. But obtaining multi-angle projections is not always feasible and complicates and extends the acquisition time. Combined corrections for both limited-view and partial volume artifacts will be incorporated in future studies to further improve the resolution and accuracy of reconstructions on scanners of this type.

\section{Conclusion}
In conclusion, we propose a deep-learning method for fast and robust PVC for cardiac SPECT imaging. The proposed network overcomes the limitations of the anatomical-guided PVC method. The network is featured with a densely-connected multi-dimension dynamic convolutions that allow the network to have adaptive convolutional kernels for each input image volume, even after the network is fully trained. The proposed network produced images with statistically comparable IMBV measurements to the gold-standard iY-PVC method, which demonstrates a strong potential for clinical implementations. However, the results presented in this paper are limited to canine studies. Further analysis is needed to evaluate the network performance in human studies for the diagnosis of CMVD. In our future studies, we also plan to validate the proposed method on other perfusion imaging tracers.

\bibliographystyle{ieeetr}
\bibliography{main.bib}

\end{document}